\documentclass[letterpaper, 10 pt, conference]{ieeeconf}  

\IEEEoverridecommandlockouts                              

\usepackage{hyperref}
\usepackage{amsmath}
\usepackage{amssymb}
\usepackage{cite}
\usepackage{graphicx}
\usepackage{float}
 \usepackage{algorithm}
\usepackage{algpseudocode}
\graphicspath{ {./images/} }
\usepackage{comment}

\title{
Computationally Efficient Data-Driven Discovery and Linear Representation of Nonlinear Systems For Control
}

\author{Madhur Tiwari,$^{1}$ George Nehma$^{1}$ and  Bethany Lusch$^{2}$
\thanks{This research was funded in part by and used resources of the Argonne Leadership Computing Facility, which is a DOE Office of Science User Facility supported under Contract DE-AC02-06CH11357.}
\thanks{$^{1}$Madhur Tiwari and George Nehma are with the Department of Aerospace, Physics and Space Sciences, Florida Institute of Technology, 150 W University Blvd, Melbourne, FL, 32901 USA
        ({\tt\small mtiwari@fit.edu, gnehma2020@my.fit.edu})}%
\thanks{$^{2}$ Bethany Lusch is a computer scientist with Argonne Leadership Computing Facility, Argonne National Laboratory, 9700 S Cass Ave, Lemont, IL 60439 USA
        ({\tt\small blusch@anl.gov})}
}

\begin{document}

\maketitle
\thispagestyle{empty}
\pagestyle{empty}

\begin{abstract}
This work focuses on developing a data-driven framework using Koopman operator theory for system identification and linearization of nonlinear systems for control. Our proposed method presents a deep learning framework with recursive learning. The resulting linear system is controlled using a linear quadratic control. An illustrative example using a pendulum system is presented with simulations on noisy data. We show that our proposed method is trained more efficiently and is more accurate than an autoencoder baseline.  
\end{abstract}
\section{Introduction}
\label{sec:I}
Linear dynamics are desirable for control due to the applicability of a rigorous control system toolkit to guarantee controllability, observability, and stability for dynamical systems.
However, nearly all dynamical systems are inherently nonlinear and thus require linearization techniques or complex nonlinear state estimation and control. Traditional linearization techniques linearize around an operating point, only approximate for small time horizons, and require real-time or recursive techniques, which increase the computational burden and unpredictability of the system. Additionally, nonlinear state estimation and control techniques often cannot guarantee stability without making several assumptions. 

Koopman theory, first proposed in 1931 \cite{Hamiltonian_Koopman}, has gained traction over the last few years as a solution for linearizing nonlinear systems. In a nutshell, the theory states that the dynamics of a system can be described linearly by an infinite-dimensional Koopman operator; due to its infinite dimensions, for practical use, it is typically approximated using data-driven methods such as Extended Dynamic Mode Decomposition (EDMD) \cite{Korda_2018,Brunton2016-cp,Lusch_2018}. Constructing a feed-forward neural network (NN) with the EDMD makes it possible to find an approximate general Koopman operator linearization applicable to a region of state space. Another advantage of using a data-driven method is that it does not require system knowledge, and the nonlinear system can be completely unknown. 

Recently, several works have focused on developing data-driven frameworks for linearization and system identification of nonlinear systems. Notably, Lusch et al. \cite{Lusch_2018} presented a data-driven method for discovering Koopman eigenfunctions. A modified autoencoder was implemented for identifying nonlinear coordinates on which the dynamics are globally linear. However, the framework handles the continuous spectra with a generalization of Koopman representations, which makes the application of linear control non-trivial. Additionally, even though the models can accurately predict longer intervals than traditional linearization schemes, error propagation still compounds. Junker et al. \cite{Junker_2022} implemented the prediction step from \cite{Korda_2018} that reduced the propagation of the error by reevaluating the observable function at every time step through the extraction of the state vector. The authors applied their implementation to a simplified golf robot and produced highly accurate representations that strongly align with the nonlinear dynamics of the robot. They showed that the technique could correctly represent system properties such as stability, controllability, and observability.

In training any neural network, the amount of data that can be generated or gathered is an important factor in how successful the network will be in its prediction. \cite{MPC_Koopman} utilizes a deep neural network (DNN) to implement a real-time, online MPC in robotic simulations. Although proving successful in implementation, some shortcomings for some simulations were large training data requirements and the handling of non-Lipschitz terms. Gathering these large amounts of data, especially in real-world scenarios, could be lengthy and difficult, as seen in \cite{VehiclePlatoons,ApplicationsSurvey, LaneKeeping}. 

In \cite{Koopman_auto}, Xiao et al. investigated using a deep learning-based EDMD approach to constructing a set of observables able to linearize vehicle dynamics for an autonomous control approach. Compared to other neural network approaches, which often lack interpretability, this method proved to be significantly better at long-term prediction due to its novel, multi-step prediction loss function, similar to the loss function we implemented in this work. A similar example to the problem presented in this paper is the inverted pendulum case. In \cite{InvertedPendulum}, the authors looked at inverting the pendulum on a cart and then comparing a traditional linearized model to that of a $4\times 4$ and $16 \times 16$-sized Koopman operator. Finding little improvement in the approximation with the larger, higher-dimensional operator was an interesting finding. The operators presented in \cite{InvertedPendulum} performed better than traditional linearization techniques for large-angle initial conditions for the pendulum. Zinage et al. \cite{Zinage2023-vz} developed a learning-based controller using Lyapunov theory. The framework ensures stability. However, system knowledge is required for physics-inspired learning, and the autoencoder needs a large amount of training data.


To this end, the contributions we propose are three-fold: (1) an implementation of a feed-forward NN model for recursive learning of deep-Koopman (RLDK) that removes the necessity for a decoder, therefore improving computation speed, reducing complexity in the model and enhancing the efficiency for real-world use, (2) an implementation of a linear quadratic regulator (LQR) controller for the system to demonstrate the ability of the learned system to be controlled effectively, and  (3) a demonstration that the RLDK learning framework is inherently robust to noisy training data as shown in later sections.  

Section \hyperref[sec:II]{II} provides background information regarding the Koopman operator and its derivation, EDMD and how it is used in the formulation of the Koopman operator, and finally, the neural network architecture used in our work. The pendulum problem used to verify this method is presented in Section \hyperref[sec:III]{III}, followed by results and discussions. Section \hyperref[sec:IV]{IV} concludes this paper.

\section{Background and Theory}
\label{sec:II}

\subsection{Koopman Operator Theory}

Koopman operator theory defines a way to transform any nonlinear dynamical system into an infinite-dimensional linear system \cite{Hamiltonian_Koopman}. Suppose we have an uncontrolled discrete-time nonlinear dynamical system defined as

\begin{equation}
    x_{k+1} = \boldsymbol{f}(x_k),
\end{equation}

where \(x_k \in \mathcal{M} \subset \mathbb{R}^n \) is the system state, \(k\) is the time index, and \(\boldsymbol{f}\) is the function that evolves the states through state space. We then define observables (unrelated to observability from control theory), which here are real-valued functions of the system state: \(g : \mathbb{R}^n \rightarrow \mathbb{R}\).  The Koopman operator, \(\mathcal{K}\), is defined such that, for any observable function $g$,
\begin{equation}
    \mathcal{K}g = g\circ \boldsymbol{f},
\end{equation}

where \(\circ\) is the composition operator. We can now apply this operator to the discrete-time system defined previously to arrive at:

\begin{equation}
    \mathcal{K}g(x_k) = g(\boldsymbol{f}(x_k)) = g(x_{k+1}).
\end{equation}

Equation (3) shows that the Koopman operator propagates an observable function of any state, \(g(x_k)\), to the next time step. 

It is not practical to apply the infinite-dimensional operator, so several methods have been developed to select a useful finite-dimensional Koopman observable space \cite{Mezic2005-py,Brunton2016-cp,Lusch_2018,Koopman_auto}. Since it is non-trivial to select a set of observable functions that span a Koopman invariant subspace, we use EDMD \cite{Williams2015-sr} to find finite-dimensional approximations of the Koopman operator. EDMD involves lifting time-series data $x_1, \dots, x_M$ into higher-dimensional space by selecting a set of observable functions and applying them to each $x_k$. Then, a linear matrix is fit to the higher-dimensional system. It is common to consider observable functions such as higher-order polynomials, radial basis functions, trigonometric functions, etc. \cite{Mezic2005-py,Folkestad_undated-tv}. Note: in this work, we include the identity function $g(x) = x$ as an observable function (see section II C). We refer readers to Brunton et al. \cite{brunton2021modern} for an in-depth study of Koopman operator theory.  

\subsection{Koopman with Control}

In this work, deep neural networks (DNNs) are employed in conjunction with EDMD to approximate the Koopman operator in finite dimensions. First, rather than hand-selecting a set of observable functions, the DNN defines a set of observable functions. Then, the finite-dimensional approximation to the Koopman operator is calculated using least-squares regression (see Algorithm \hyperref[alg:1]{1}).

\begin{figure*}
    \centering
    \includegraphics[height=2.6in]{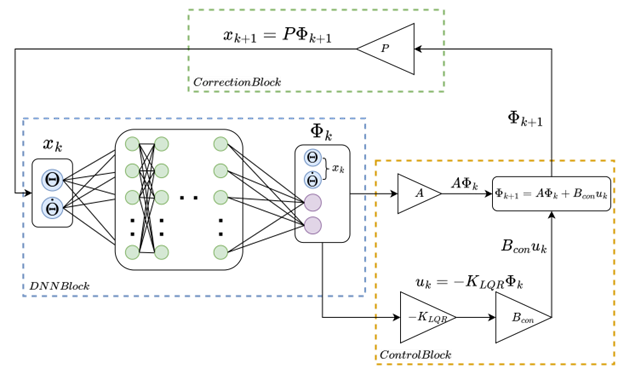}
    \caption{Complete schematic of dynamic propagation of the states, including the control input. Note that a decoder is not needed.}
    \label{fig:1}
\end{figure*}

For a controlled system, the nonlinear dynamics are propagated with the following mathematical expression:

\begin{equation}
    x_{k+1} = \boldsymbol{f}(x_k,u_k),
\end{equation}

where \(x_k \in \mathbb{R}^n \) and \(u_k \in \mathbb{R}^p \). After lifting the states to higher dimensions with $\boldsymbol{\Phi}$, we wish to find a linear input-output system representation

\begin{equation}
    \mathbf{\Phi}(x_{k+1}) \approx \mathbf{K} \boldsymbol{\Phi}(x_k) + \mathbf{B}\mathbf{u}_k,
\end{equation}

where the matrix $\mathbf{K}$ approximates the Koopman operator. This is analogous to $\mathbf{A}$ and $\mathbf{B}$, the state transition and input matrices, respectively, in linear control systems. The goal of the deep learning framework (Fig. \ref{fig:1}) is to learn \(N\) observables/lifting functions and calculate the $\mathbf{K}$ and $\mathbf{B}$ matrices. We choose \(N\)  such that \(N >> n\) and define

\begin{equation}\label{eqn:observ}
    \mathbf{\Phi}(\textbf{x}_k) := 
    \begin{bmatrix} 
    \textbf{x}_k \\
    \phi_1(\textbf{x}_k) \\ \phi_2(\textbf{x}_k) \\ \vdots \\ \phi_N(\textbf{x}_k) 
    \end{bmatrix},
\end{equation}

where \(\mathbf{\phi}_i : \mathbb{R}^n \rightarrow \mathbb{R}, i = 1,...,N\) are the observable functions, which are defined by the DNN. 
Note that we concatenate the original states \(\mathbf{x}_k\) with the observables from the DNN. This helps in two ways: (1) the original states can be easily extracted from the observables for control (see below), and (2) there is no need for a decoder to extract the original, thus saving valuable computational resources.
Currently, there is no method to dictate the size of \(N\) that would guarantee the best balance between simplicity and accurate approximation of the system; therefore, in most cases, \(N\) is chosen empirically through trial and error. 
Some work has been done to study the size $N$ to find controllable systems \cite{Zinage2023-vz,Lusch_2018}. In this work, DNNs are applied with EDMD to find the Koopman operator.  

To calculate the approximate Koopman operator $\mathbf{K}$ and the input matrix $\mathbf{B}$, the time history of measurement data for \(M\) steps is arranged into snapshot matrices. The first matrix, \(X\), is the state history from time \(k=1\) to \(k=M-1\), whilst the second matrix, \(X'\) is the same state history, right shifted by one-time step: 

\begin{equation}
    X = \begin{bmatrix} x_{1},  x_{2},  x_{3},  \dots, x_{M-1} \end{bmatrix} 
\end{equation}
\begin{equation}
    X' = \begin{bmatrix} x_{2},  x_{3}, x_{4},  \dots, x_{M} \end{bmatrix} 
\end{equation}

Mapping the measured states with observable functions leads to

\begin{equation}
    \boldsymbol{\Phi}(X) = \begin{bmatrix} \boldsymbol{\Phi} (x_1), \boldsymbol{\Phi}(x_2),  \dots, \boldsymbol{\Phi}(x_{M-1}) \end{bmatrix} 
\end{equation}
\begin{equation}
    \boldsymbol{\Phi}(X') = \begin{bmatrix} \boldsymbol{\Phi}(x_2), \boldsymbol{\Phi}(x_3),  \dots, \boldsymbol{\Phi}(x_M)  \end{bmatrix} 
\end{equation}    

Given the dataset, the matrices $\mathbf{K}$ and $\mathbf{B}$ can be found using least-sqaures

\begin{equation}
\min \sum\left\|\boldsymbol{\Phi}(x_{k+1})-\left(\mathbf{K} \mathbf{\Phi}\left(x_k\right)+\mathbf{B} u_k\right)\right\|^2.
\end{equation}


Applying the snapshot matrices of real data yields

\begin{equation}
    \mathbf{\Phi(X')} \approx 
    \mathbf{K} \mathbf{\Phi(X)}+ \mathbf{B}\mathbf{U} = \begin{bmatrix}
        \mathbf{K} & \mathbf{B}
    \end{bmatrix}
    \begin{bmatrix}
        \mathbf{\Phi(X)} \\ \mathbf{U}
    \end{bmatrix};
\end{equation}

therefore, 

\begin{equation}
    \begin{bmatrix}
        \mathbf{K} & \mathbf{B}
    \end{bmatrix} = \boldsymbol{\Phi(X')}\begin{bmatrix}
        \mathbf{\Phi(X)} \\ \mathbf{U}
    \end{bmatrix}^\dagger,
\end{equation}

where $\dagger$ denotes the Moore-Penrose inverse of the matrix \cite{penrose_1955}. 

Because the Koopman operator calculated with least-squares is an approximation, as the observable functions do not span a Koopman invariant subspace, the predicted state is an approximation, which we denote with the $\hat{}$ symbol:

\begin{equation}
    \hat{\mathbf{\Phi}}(x_{k+1}) = \mathbf{K} \hat{\mathbf{\Phi}}(x_k) + \mathbf{B} \mathbf{u}_k.
\end{equation}



Now, we can extract the original states from the observables using a projection matrix $\mathbf{P}$ \cite{Junker_2022} yielding

\begin{equation} \label{eqn:extract}
    x_{k+1} = \mathbf{P}\hat{\mathbf{\Phi}}(x_{k+1})\; \text{with}\; \mathbf{P} = \begin{bmatrix}
        \mathbf{I}_n, \mathbf{0}_{n \text{x} N}
    \end{bmatrix},
\end{equation}

where \(\mathbf{I}_n\) is the \(n\) x \(n\) identity matrix and \(\mathbf{0}_{n \text{x} N}\) is the \(n\; \text{x}\; N\) zero matrix. As shown in \cite{Junker_2022} and \cite{Korda_2018}, the observable functions not spanning a Koopman invariant subspace leads to error accumulation over time, which can lead to misleading predictions. However, if the prediction is corrected at each time step, then this error can be mitigated. This correction is applied by extracting the estimated state variable $\hat{x}_{k+1}$ at each time step with Equation \ref{eqn:extract} and then reapplying the observables to the extracted state variable with Equation \ref{eqn:observ}. 

\subsection{Neural Network Architecture}

One of the key improvements introduced in this paper is the use of a neural network with a recursive learning framework for learning the Koopman matrix to improve the training speed and reduce the time to predict the next step. Ultimately, more complex dynamical systems would inherently require more complex, larger Koopman operators in a higher dimensional lifted space, and the proposed architecture could facilitate that case. 

As a baseline, we compare to an autoencoder, which is a type of NN architecture in which the output is the same as the input with an intermediate state in the middle that is ``encoded.'' Some of the most common uses of an autoencoder are in image compression, anomaly detection, and data denoising \cite{LI}, where the encoder transforms the original data into the (generally compressed) encoded space, and the decoder reconstructs the original data from that encoded space. To use the autoencoder architecture with the Koopman operator, we encode the states into a higher dimensional lifted space with the encoder. One purpose of an autoencoder for Koopman operator theory is to recover the original states for control uses, which, in our case, is bypassed by concatenating the original states, as shown earlier. 

Figure \ref{fig:1} shows the proposed architecture for the trained network with prediction and control in continuous time. Here, the given initial states pass through the DNN block to form the lifted states. The lifted states are then concatenated with original states $\mathbf{x}_k$ to form the new set of observables $\mathbf{\Phi}$. The states are propagated through the control block where $\mathbf{A}$ and $\mathbf{B}_{con}$ are the continuous form of Koopman matrix $\mathbf{K}$ and $\mathbf{B}$ (see section III B). The predicted states then pass through the correction block, resulting in the original states' extraction. The entire loop runs until the states are regulated. 

The complete algorithm can be seen in Algorithm 1, and the full PyTorch code is on Github\footnote[1]{\url{https://github.com/tiwari-research-group/Koopman-Control-No-Decoder}}.

The NN is trained using traditional batch training with a custom loss function. The loss is calculated recursively by applying the mean squared error (MSE) and EDMD. The prediction loss function is given as 

\begin{equation}
\mathcal{L}_{\text {pred}}=\frac{1}{N_d} \sum_{k=1}^{N_d}\left\|\hat{\boldsymbol{\Phi}}({x_{k+1}})-\mathbf{\Phi}({x_{k+1}}) \right\|_2^2.
\end{equation}

Here $\hat{\boldsymbol{\Phi}}({x_{k+1}})$ is the predicted observable defined in Equation 14, where $\mathbf{K}$ and $\mathbf{B}$ are calculated recursively using the least-squares method as in Equation 13, and  $\mathbf{\Phi}({x_{k+1}})$ are the observables calculated using the true data.  

The data generation pipeline for the batch training can be seen in Algorithm 2. It can be noted that to collect data, the system is excited using random control input (+1 or -1). Also, random Gaussian noise is added to the training data to test the effectiveness of the proposed framework.

\begin{algorithm}
\caption{Learning Koopman and Input Matrix}\label{alg:1}
\begin{algorithmic}[1]
\Statex \textbf{Input: } \text{X,Y,u training data, batch size \(b_s,\) Epoch\(_{max}\)}
\Statex \textbf{Method}
\For{\text{epoch in range(Epoch\(_{max}\))}}
\For{\text{batch in range(number of batches)}}
\State \text{Sample \(b_s\) from training, control and label data}

\Statex \hspace{\algorithmicindent} \hspace{\algorithmicindent} \text{\(X = \{X^i_{0:k-1}\}^{b_s}_{i=1}, Y = \{Y^i_{1:k}\}^{b_s}_{i=1}\)}
\Statex \hspace{\algorithmicindent} \hspace{\algorithmicindent} \text{\(u = \{u^i_{0:k-1}\}^{b_s}_{i=1}\)}
\State \text{Encode the training and label data and stack the}

\Statex \hspace{\algorithmicindent} \hspace{\algorithmicindent} \text{original states onto the encoded states.}

\Statex \hspace{\algorithmicindent} \hspace{\algorithmicindent} \text{\(\Phi_x = [X; \Phi(X)]^T, \Phi_y = [Y; \Phi(Y)]^T\)}

\State \text{Compute the W,V matrices and their transposes.}

\Statex \hspace{\algorithmicindent} \hspace{\algorithmicindent} \text{\(W = [\Phi_x;u], V = \Phi_y\)}

\State \text{Use Least Squares to find the KB matrix and}

\Statex \hspace{\algorithmicindent} \hspace{\algorithmicindent} \text{extract K and B from KB matrix.}

\Statex \hspace{\algorithmicindent} \hspace{\algorithmicindent} \text{\(KB = VW^T \cdot (WW^T)^\dagger\)}
\State \text{Compute the next state with the linear equation.}

\Statex \hspace{\algorithmicindent} \hspace{\algorithmicindent} \text{\(\hat{\Phi}_{x+1} = K\Phi_x + Bu\)}
\State \text{Apply the loss function \(\mathcal{L}\) using MSE}
\State \text{Update the model weights and parameters with}
\Statex \hspace{\algorithmicindent} \hspace{\algorithmicindent} \text{the Adam optimizer.}
\EndFor
\EndFor
\Statex \textbf{Output}: \text{K and B matrices, trained DNN}
\end{algorithmic}
\end{algorithm}

\subsection{LQR Controller}

An LQR controller is employed with the learned dynamics. The objective of the LQR controller is to regulate the states of a system through the use of a quadratic loss function. The loss function \(\mathcal{J} \) is given by

\begin{equation}
    \mathcal{J} = \int_{0}^{\infty} \frac{1}{2}[\mathbf{\Phi}(x)^T\mathbf{Q\Phi}(x) + \mathbf{u}^T\mathbf{Ru}]dt,
\end{equation}

where $\mathbf{Q} = \begin{bmatrix} I_{n \times n} &. & .& 0\\. &. & .& 0\\. &. & .& 0\\0 &0 & 0& 0 \end{bmatrix} \in R^{N \times N}$ and $\mathbf{R} = 1$ are the tuning matrices for states and control respectively. It can be noted that since the first $n$ rows of the observable $\mathbf{\Phi}(x)$ are the states themselves, it is straightforward to weight the controllable states. The LQR control effort is given as  

\begin{equation}
    u(t) = -K_{LQR}\mathbf{\Phi}(x).
\end{equation}

This constant gain is found in MATLAB by calling the \emph{LQR} function with the appropriate input variables. The trajectory of the LTI system can then easily be found by integrating it with the RK4 function. For purposes of control, we converted the Koopman matrix $\mathbf{K}$ and input matrix $\mathbf{B}$ in continuous form such that 

\begin{equation}
\mathbf{A} = \frac{\mathbf{K-I}}{\Delta t} \ \ \text{and} \ \ \mathbf{B_{con}} = \frac{\mathbf{B}}{\Delta t},
\end{equation}

where $\Delta t$ is the RK4 time step of $0.01$ seconds.

\section{Simulation and Results}
\label{sec:III}

In this section, we present the simulation, results, and discussion of the proposed method. A comparison of the proposed methodology against a traditional architecture with an autoencoder-based framework is also presented. 

\subsection{Simulation Setup}

To demonstrate the applicability of our model, we have chosen to use the pendulum problem. It is a common example that is widely used due to its relatively simple yet nonlinear dynamics. 
The nonlinear state dynamics are given as follows:

\begin{equation}
    \begin{bmatrix} x_1 \\ x_2
    \end{bmatrix} = 
    \begin{bmatrix} \theta \\ \Dot{\theta} 
    \end{bmatrix}
    \label{pend 1}
\end{equation}

\begin{equation}
    \Dot{x} = \begin{bmatrix} \Dot{x_1} \\ \Dot{x_2}
    \end{bmatrix} = 
    \begin{bmatrix} x_2 \\ -\frac{g}{l}sin(x_1) 
    \end{bmatrix}
    \label{pend 2}
\end{equation}

To generate training data, the pendulum dynamics were integrated for 2 seconds using a Runge-Kutta integrator, using random initial conditions within \(-2 \leq \theta_0 \leq 2\). For each initial condition, two data sets were created to provide the inputs and labels for the custom loss function. The first set, \(X\) was the state history from time \(k=0\) to \(k=M-1\), whilst the second set, \(X'\), was the same state history, right shifted by one time step: 

\begin{align*}
    X = \begin{bmatrix} x_{0},  x_{1},  x_{2},  \dots, x_{M-1} \end{bmatrix} 
\end{align*}
\begin{align*}
    X' = \begin{bmatrix} x_{1},  x_{2}, x_{3},  \dots, x_{M} \end{bmatrix}. 
\end{align*}

The training data is comprised of 8000 different random initial conditions. From this, 80\% of the data is used in the training set, whilst 10\% is used for the validation set and 10\% is used as the test set. The duration for each initial condition is 2 seconds with a \(\Delta t=0.01\).

\begin{algorithm}
\caption{Data Generation for Noisy Data w/ Control}\label{alg:cap}
\begin{algorithmic}[1]
\Statex \textbf{Input: } \text{Number of Initial Conditions, time step \(dt\),}
\Statex \text{final time \(t_f\), data points \(dp\), state length \(n_x\),}
\Statex \text{control input length \(m\).}
\Statex \textbf{Method}
\For{\text{num in range(number of Initial Conditions)}}

\State \text{Generate random I.C with noise}

\Statex \hspace{\algorithmicindent} \text{\(x_{init} = 2* torch.Tensor(2).uniform\_(-1,1) +\)} 
\Statex \hspace{\algorithmicindent} \text{\((0.0001**0.5)*np.random.normal(0,1,2)\)}
\State \text{Declare the time vector for each I.C}

\Statex \hspace{\algorithmicindent} \text{\(time = torch.linspace(0,t_f,dp)\)}

\For {\text{t in time}}
\State \text{Generate random control input}

\Statex \hspace{\algorithmicindent} \hspace{\algorithmicindent} \text{\(u_k = torch.Tensor(1).uniform\_(-1,1)\)}

\State \text{Save the current state in the X data array}

\State \text{Integrate the current state using the dynamics}
\Statex \hspace{\algorithmicindent} \hspace{\algorithmicindent}\text{and RK4 functions. Add noise to new state}

\Statex \hspace{\algorithmicindent} \hspace{\algorithmicindent} \text{\(x_{k+1} = rk4(pendfunc,t,x_k,u_k,dt)+\)}
\Statex \hspace{\algorithmicindent} \hspace{\algorithmicindent} \text{\((0.0001**0.5) *np.random.normal(0,1,2)\)}
\State \text{Save the new state in the Y data array.}
\EndFor
\State \text{Stack the solutions, X,Y,u from the I.C. into 3D}
\Statex \hspace{\algorithmicindent} \text{array that holds every trajectory for each I.C.}

\Statex \hspace{\algorithmicindent} \text{\(data_i = torch.vstack([data_i,new_i[None,:]])\)}
\EndFor

\Statex \textbf{Output}: \text{Data matrices, \(data_x\), \(data_y\), \(data_u\),}
\Statex \hspace{\algorithmicindent} \hspace{\algorithmicindent} \text{input parameters.}
\end{algorithmic}
\end{algorithm}

The training data includes a random control effort of -1 or 1 to simulate the control effort that would be normally applied when deployed. This randomly excites the system and enables the extraction of the control input matrix $\mathbf{B}$. Additionally, to check the robustness of the model, random Gaussian noise is added to each state to simulate sensor measurement noise. This is achieved by adding noise with the same standard deviation to each state at every integration time step with \(\mu = 0.0158\)  and \(\sigma = 0.00913\). See Algorithm \ref{alg:cap}.

\subsection{Results and Discussions}
\label{sec:IV}

\begin{figure}[hbt]
    \centering
    \includegraphics[width=0.48\textwidth]{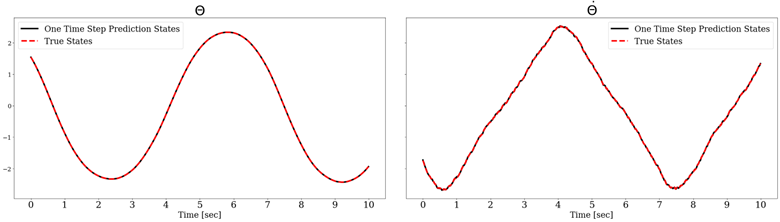}
    \caption{Learned Koopman Operator dynamics prediction for only one time step in the future with comparison to the ground truth nonlinear dynamics.}
    \label{fig:enter-one time setp}
\end{figure}

\begin{figure}[hbt]
    \centering
    \includegraphics[width=0.48\textwidth]{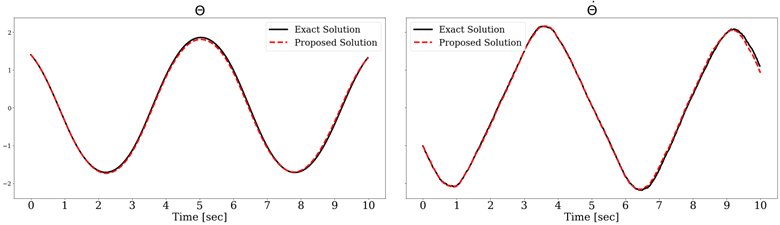}
    \caption{Learned, self-propagating dynamics prediction given only the same initial condition as the ground truth nonlinear dynamics}
    \label{fig:enter-10 seconds}
\end{figure}

\begin{figure}[hbt]
    \centering
    \includegraphics[width=0.48\textwidth]{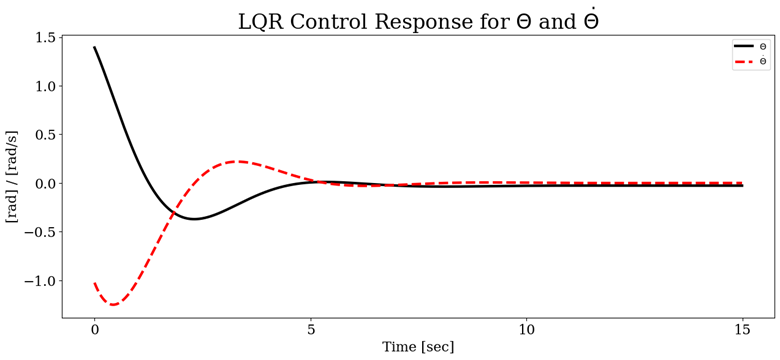}
    \caption{System response to LQR Controller developed using the linear system created with the learned Koopman operator.}
    \label{fig:LQR}
\end{figure}

\begin{figure}[hbt]
    \centering
    \includegraphics[width=0.48\textwidth]{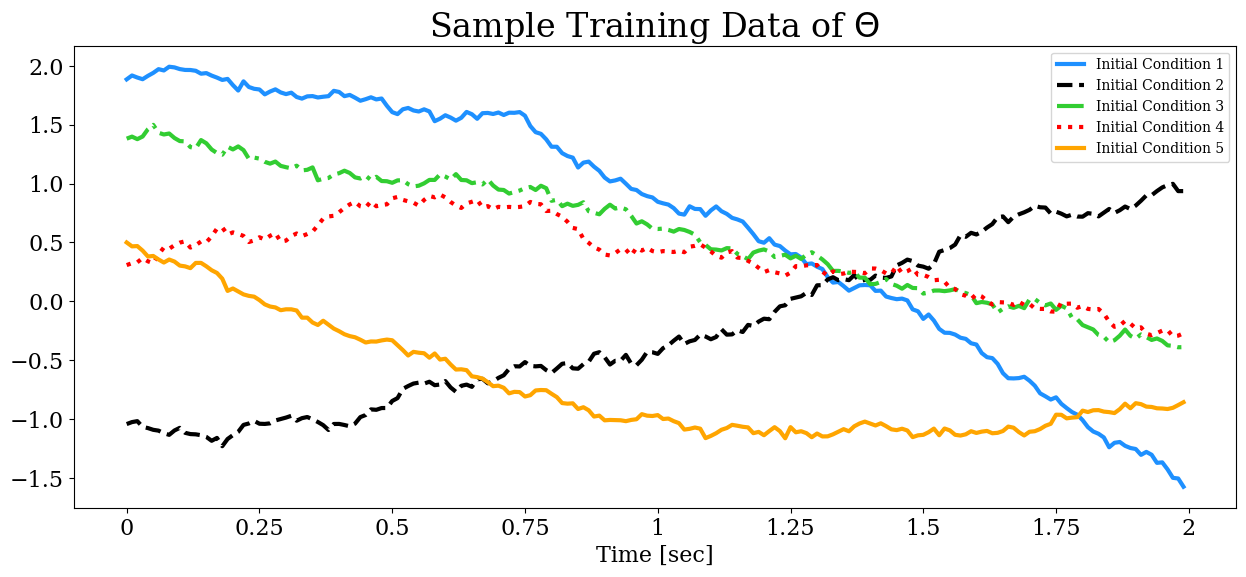}
    \caption{Five random trajectories from the training set, highlighting the noise in the data.}
    \label{fig:training data}
\end{figure}

\begin{figure}[hbt]
    \centering
    \includegraphics[width=0.48\textwidth]{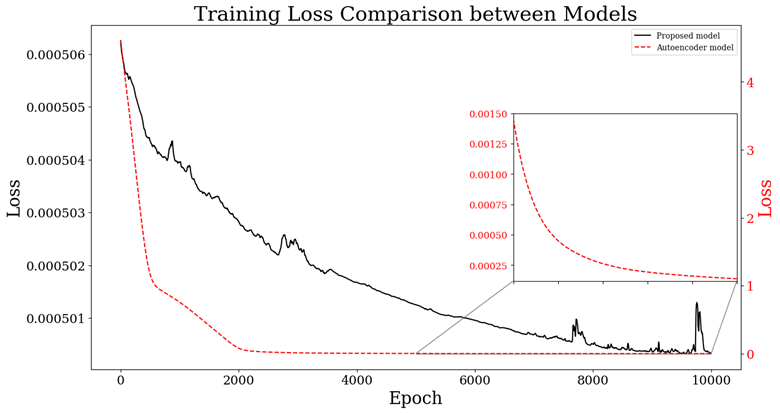}
    \caption{Training loss comparison between the autoencoder model and our proposed model. Our model immediately begins with a lower loss due to weight initialisation.}
    \label{fig:training error}
\end{figure}

In this subsection, we discuss the results of the proposed architecture. We also compare the performance of the proposed architecture with a traditional learning framework with an autoencoder. 
Figure \ref{fig:enter-one time setp}  shows the prediction for only one step compared to the true dynamics. Figure \ref{fig:enter-10 seconds} shows the prediction for 10 seconds, given only the initial states. It can be seen that learned dynamics closely follow true nonlinear dynamics. However, beyond 10 seconds, there is a significant divergence from the true dynamics, highlighting the limitations of the finite representation of the Koopman operator. The system response when subjected to LQR control is shown in Figure \ref{fig:LQR}. It can be noted that other controllers such as a model predictive control can be employed in this case. Evident in Figure \ref{fig:training data}, the data that is used for the training of the DNN  includes noise to mimic the inputs from a real-world sensor.  As shown in the full prediction in Figure \ref{fig:enter-10 seconds}, the DNN learns to predict smoother dynamics than the noisy training data, shown in Figure \ref{fig:training data}. It can be seen in Figure \ref{fig:training error} that the loss of both models observes the anticipated downward trend.

\begin{figure}[hbt]
    \centering
    \includegraphics[width=0.48\textwidth]{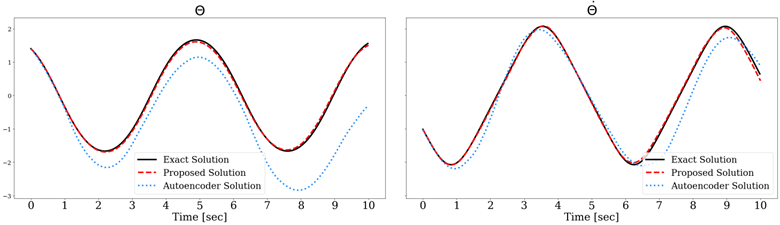}
    \caption{Comparison of the learned dynamics between the model with autoencoder and the new model.}
    \label{fig:comparison}
\end{figure}

\begin{figure}[hbt]
    \centering
    \includegraphics[width=0.48\textwidth]{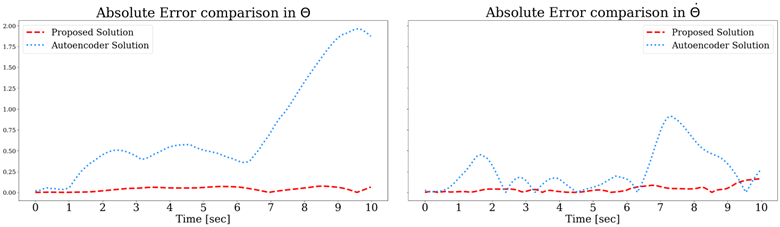}
    \caption{Absolute Error (difference between prediction and truth) for our proposed solution and the autoencoder.}
    \label{fig:error comparison}
\end{figure}

Figure \ref{fig:comparison} shows the 10 second comparison between the true dynamics, the proposed method, and the traditional autoencoder. It can be seen that the proposed model more closely follows the true dynamics. Figure \ref{fig:error comparison} shows the absolute error between the true dynamics, the proposed method, and the traditional autoencoder solution. It can be noted that the proposed method is more accurate than the autoencoder framework. Additionally, the training time of our method was 30 seconds, and autoencoder training took 54 seconds. Therefore, the proposed method trains more efficiently than the traditional framework.

\section{Conclusions}
\label{sec:VI}
In this work, an updated framework for learning the Koopman operator with control is proposed. It was found that the proposed method is more computationally efficient and accurate than the baseline Koopman learning architecture with an autoencoder. The proposed method is tested in simulation on a pendulum problem with LQR control, and it was seen that the method is able to closely follow the true nonlinear dynamics for at least 10 seconds.
Future work involves testing and verifying the framework for complex systems with experimental data. Furthermore, ensuring the stability of the learning and control with further testing against noisy and impartial data is another topic of future study. 





\bibliographystyle{ieeetr}
\bibliography{bibliography.bib}

\end{document}